\def \pom {{\hspace{ -0.1em}I\hspace{-0.2em}P}}

\documentstyle[12pt,psfig]{article}
\begin{document}
\begin{titlepage}
\begin{center}

{\Large\bf{ Hard and Soft Contributions in  Diffraction: A Closer Look}}
\\[5.0ex]
{ \Large \it{ M.B. Gay Ducati
$^{1,\dag}$\footnotetext{$^{\dag}$E-mail:gay@if.ufrgs.br}, V.P.
Gon\c{c}alves $^{2,*}$\footnotetext{$^{*}$E-mail:barros@ufpel.tche.br} and
M.V.T. Machado
$^{1,\star}$\footnotetext{$^{\star}$E-mail:magnus@if.ufrgs.br}}}
 \\[1.5ex]
{\it $^1$  Instituto de F\'{\i}sica, Universidade Federal do Rio Grande do
Sul\\ Caixa Postal 15051, CEP 91501-970, Porto Alegre, RS, BRAZIL}\\
{\it $^2$  Instituto  de F\'{\i}sica e Matem\'atica, Universidade
Federal de Pelotas\\ Caixa Postal 354, CEP 96010-090, Pelotas, RS, BRAZIL}
\end{center}
{\large \bf Abstract:}
Disentangle  the hard and soft dynamics in diffractive DIS is one of the main
open questions of the strong interactions. We propose the study of the 
logarithmic slope  in $Q^2$ of the diffractive structure function as a
potential observable to discriminate between the Regge and the QCD-based
approaches. Our results indicate that a future experimental analyzes could
evidentiate the leading dynamics at $ep$ diffractive processes in the HERA
kinematical regime.

\vspace{1.5cm}



\end{titlepage}

The study of  electroproduction at small $x$ has lead to the improvement of
our understanding of QCD dynamics at the interface of perturbative and
nonperturbative physics. However, many important problems  remain at present
unsolved. A longstanding puzzle in the particle physics is the nature of the
Pomeron. This object, with the quantum numbers of the vacuum, was introduced
phenomenologically in the Regge theory  as a simple moving pole in the complex
angular momentum  plane, to describe the high-energy behavior of the total and
elastic cross-sections of the hadronic reactions \cite{collins}. Within the
framework of the perturbative  QCD (pQCD), the Pomeron is associated with the
resummation of leading logarithms in $s$ (center of mass  energy squared) and
at  lowest order is described by the two-gluon exchange \cite{DDIS}. Due to
its zero color charge the Pomeron is associated with diffractive events, 
characterized by the presence of large rapidity gaps in the hadronic final
state,  which are exponentially suppressed \cite{bjorken}. Diffractive
processes in deep-inelastic scattering (DIS) are of particular interest,
because the hard photon in the initial state gives rise to  the hope that, at
least in part, the scattering amplitude can be calculated in pQCD. Moreover,
DIS exhibits the nice feature of having a colorless particle, the virtual
photon, in the initial state. The main theoretical  interest in diffraction is
centred around the interplay between the soft and hard physics. Hard physics
is associated with the well established parton picture and perturbative QCD,
and is applicable to processes for which a large scale is present. Soft
dynamics on the other hand, linked for example with the total cross section 
of hadron scattering, is described by nonperturbative aspects of QCD. The
ability to separate clearly the regimes dominated by soft and hard processes
is essential in exploring QCD at both  quantitative and  qualitative level.

In  DIS   the partonic fluctuations of the virtual photon  can lead to
configurations of different sizes when analysed in the proton rest frame. The
size of the configuration will depend on the relative transverse momentum
$k_T$ of the $q\overline{q}$ pair. The small size configurations are
calculated using perturbative QCD and at small Bjorken scaling variable $x$
(large $s$) the smallness of the cross section (color transparency) is
compensated by the large gluon distribution. For large size configurations one
expects to be in the regime of soft interactions. In the inclusive measurement
of diffractive final states, where the diffractive structure function is
derived, one sums over both small-distance and large-distance configurations.
So far there is no theoretical framework which allows one to predict the
relative magnitudes of the "soft" and  the "hard" components of the
diffractive cross section. One possibility is the analyzes of the energy
dependence of the cross section, since we expect that the "soft" component
rises weakly with the energy for any fixed mass of the diffractive system,
whereas the "hard" part should rise faster. As in the diffractive cross
section we integrate over both the perturbative and nonperturbative regions of
the phase space, there is a competition between these two pieces. At first
sight, the large momentum region seems to be rather subdominant. However, the
large gluon distribution function provides an enhancement in this region, and
in this way weakening the dominance  of the soft nonperturbative region. As a
result, the effective value of the exponent $n$ of the energy dependence lies
between the hard ($n_{hard}  \approx 1.4$ ) and the soft ($n_{soft}\approx
1.12$) values \cite{f2d3}.

Since the first observation of diffractive DIS at HERA, several attempts have
been made to compare the data with the Regge and QCD-based models
\cite{compara,WM}.   In general, these models provide a reasonable description
of the present data on the diffractive structure function $F_2^{D}$, although
based on quite  distinct frameworks, demonstrating the inclusive character of
this observable to delimit the interplay of soft and hard QCD in diffraction. 
In this letter we propose the analyzes of the  logarithmic slope of the
diffractive structure function as a potential observable to clarify the
dynamics in this process. Our  analyzes is  motivated by the recent
discussions in the literature about the behavior of the logarithmic slope of
the inclusive structure function $F_2(x,Q^2)$ as a possible signal of one new
regime of  QCD \cite{newregime}. At the moment  $ep$ HERA data on the $F_2$
slope cannot clearly demonstrate the presence of a new dynamics in its
kinematical regime, but new studies in $eA$ should distinguish the distinct
regimes of QCD \cite{slopevic}.  We believe that the experimental analyzes of
the logarithmic slopes of  $F_2^{D}$ will allow  to discriminate the different
contributions to the dynamics already in the current HERA kinematical region.

We study in detail  the predictions to this observable considering two distinct approaches:  i) a Regge inspired model \cite{CKMT1,CKMT2}, where the diffractive production is dominated by  a nonperturbative Pomeron, and the diffractive structure function is obtained using the Ingelman-Schlein ansatz \cite{ingelman}.
ii) a pQCD approach \cite{BEKW} where the diffractive process is modeled as the scattering of the photon Fock states with the proton through a gluon ladder exchange (in the proton rest frame). 
Before the proper analyzes of the models  we need to define the diffractive processes and the usual kinematical variables (for a review, see Ref. \cite{DDIS}). The most important observable at diffractive DIS (DDIS) is the associated structure function $F_2^D$ \cite{f2d3}. In this work we are concerned to the $t$-integrated structure function,  denoted  $F_2^{D(3)}$.
The main variables used for the description of DDIS  are the total hadronic
energy $W$ of the $\gamma^*$-proton system and the diffractive produced  mass
$M$. In the analyzes of $F_2^D$, it is convenient to use also the variables $
\beta$ and $x_{\pom}$. In terms of $W$ and $M$, one has $\beta = Q^2/(Q^2 +
M^2)$ and $x_{\pom} = (M^2 + Q^2)/(W^2 +  Q^2)$, where we have neglected the
proton mass and the momentum transfer $t$. To connect these variables with the
Bjorken scaling variable $x$, we remind  that $x = Q^2/(W^2 + Q^2)$, which
immediately leads to $x = \beta x_{\pom}$. In the kinematic domain of the
present experimental measurements, $x_{\pom}$ may be interpreted as the
fraction of the four-momentum of the proton carried by the diffractive
exchange, the Pomeron, if such a picture is invoked. The $\beta$ is the
fraction of the four-momentum of the diffractive exchange carried by the
parton interacting with the virtual boson. 

Diffraction dissociation of virtual photons, observed at HERA $ep$ collider,
furnishes the details on the nature of the Pomeron and  on its partonic
structure. Capella-Kaidalov-Merino-Tran Thanh Van (CKMT) proposed a few years
ago a model to diffractive DIS based on Regge theory \cite{CKMT1,CKMT2} and
the Ingelman-Schlein ansatz, which  is based on the intuitive picture of a
Pomeron flux associated with the proton beam and on the conventional partonic
description of the Pomeron-photon collision. In this case, deep inelastic
diffractive scattering proceeds in two steps (the Regge factorization): first
a Pomeron is emitted from the proton and then the virtual photon is absorbed
by a constituent of the Pomeron, in  the same way as the partonic structure of
the hadrons.  In the CKMT model the structure function of the Pomeron,
$F_{\pom}(\beta,Q^2)$, is associated to the deuteron structure function through
the arguments given above.  The  Pomeron is considered as a Regge pole with a
trajectory $\alpha_{\pom}(t)=\alpha_{\pom}(0) + \alpha^{\prime}\, t $
determined from soft processes, in which absorptive corrections (Regge cuts)
are taken into account. Explicitly, $\alpha_{\pom}=1.13$ and
$\alpha^{\prime}_{\pom}=0.25\;GeV^{-2}$.  The diffractive contribution to DIS
is written in the factorized form: \begin{eqnarray}
F_2^D(x,Q^2,x_{\pom},t)=\frac{[g^{\pom}_{pp}(t)]^2}{16
\pi}x_{\pom}^{1-2\alpha_{\pom}(t)}F_{\pom}(\beta,Q^2,t)\;, \end{eqnarray}
where  $g^{\pom}_{pp}(t)=g^{\pom}_{pp}(0)\,\,exp(C\,t)$ is the Pomeron-proton
coupling, with $[g^{\pom}_{pp}(0)]^2=23 \,\,mb$ and $C=2.2 \,\,GeV^{-2}$.  In
this approach, $F_{\pom}$ is determined using Regge factorization and the
values of the triple Regge couplings determined from soft diffraction data.
Namely, the Pomeron structure function is obtained from $F_2^p$, or more
precisely from the combination $F_2^d=\frac{1}{2}(F_2^p + F_2^n)$, by
replacing the Reggeon-proton couplings by the corresponding triple reggeon
couplings (see Ref. \cite{CKMT1} for details). The following parametrization
of the deuteron structure function $F_2^d$ at moderate values of $Q^2$, based
on Regge theory, was introduced: \begin{eqnarray} F_2^d(x, Q^2) & = &  A  \,
x^{- \Delta(Q^2)}(1 - x)^{n(Q^2)+4} \left ( {Q^2 \over Q^2 + a} \right )^{1 +
\Delta(Q^2)}  \nonumber \\ & + & B \  x^{1 - \alpha_R} (1 - x)^{n(Q^2)} \left
( {Q^2 \over Q^2 + b} \right )^{\alpha_R}  \end{eqnarray} where $1 +
\Delta(Q^2)$ is the Pomeron intercept, which depends on the photon virtuality,
and $\alpha_R$ is the intercept of the secondary reggeon  (the $f$ trajectory).
The Pomeron structure function $F_{\pom}$ is identical to $F_2^d$, given
above, except for the following changes in its parameters:
\begin{eqnarray}
F_{\pom}(\beta, Q^2, t) = F_2^d \left (x \to  \beta; A \to eA, B \to f B,
n(Q^2) \to n(Q^2) - 2 \right ) \,\,. 
\label{deut}
\end{eqnarray} 
 The values of  $e$ and
$f$ in $F_{\pom}$ are obtained from conventional triple reggeon fits to high
mass single diffraction dissociation for soft hadronic processes. The
remaining  parameters  are given in Refs. \cite{CKMT1,CKMT2}. 

The comparison of the CKMT model with data is quite satisfactory
\cite{CKMT1,CKMT2}. A remark is that here we use the pure CKMT model
\cite{CKMT1} rather than to include QCD evolution of the initial conditions
\cite{CKMT2}, which has been used to improve the  model at higher $Q^2$. Such
procedure ensures that we take just a pure Regge model, without contamination
from QCD inspired phenomenology.

On the other hand, the  pQCD framework has been recently used by  some
authors to describe the diffractive structure function \cite{authors}, and
their   main properties are very similar. We consider for our analyzes the
Bartels-W\" usthoff model and its further parameterization to experimental
measurements \cite{BEKW}. The physical picture is that, in the proton rest
frame, diffractive DIS is described by the interaction of the photon Fock
states ($q\bar{q}$ and $q\bar{q}g$ configurations) with the proton through a
Pomeron exchange, modeled as a two hard gluon exchange. The corresponding
structure function contains the contribution of $q\bar{q}$ production to both
the longitudinal and the transverse polarization of the incoming photon and of
the production of $q\bar{q}g$ final states from transverse photons.

The basic elements of this approach are the photon light-cone wave function
and the nonintegrated gluon distribution (or dipole cross section in the
dipole formalism). For elementary quark-antiquark final state, the wave
functions depend on the helicities of the photon and of the (anti)quark. For
the $q\bar{q}g$ system one considers a gluon dipole, where the pair forms an
effective gluon state associated in color to the emitted gluon and only the
transverse photon polarization is important. The interaction with the proton
target is modeled by two gluon exchange, where they couple in all possible
combinations to the dipole. Then the diffractive structure function can be
written as \begin{eqnarray} F_2^D(x_{\pom},\beta,Q^2) \sim
\beta\,\int\,dt\,\int\,\frac{k_t^2\,d^2k_t}{(1-\beta)^2} \,\,\left | \int \,
\frac{d^2l_t}{l_t^2}\,D\Psi(\alpha,k_t) {\cal F}(l_t^2,k_0^2;x_{\pom})\right
|^2\;, \end{eqnarray} where $D\Psi$ is a combination of the concerned wave
functions, $l_t$ is the transverse momentum of the  exchanged gluons   and
${\cal F}(l_t^2,k_0^2;x_{\pom})$ defines the Pomeron amplitude (nonintegrated
gluon distribution). The $k_0^2$ sets the hadronic scale which splits the
regions of soft and hard QCD.  With a suitable anzatz for the $l_t$ dependence
of the two-gluon Pomeron, or more precisely the non-integrated gluon
distribution, it is possible to interpolate between the hard region where the
parton model applies and the soft region where the aligned jet configuration
dominates, as emphasized in Ref. \cite{kogut}. Regarding the $x_{\pom}$
behavior, the hypothesis is that for small transverse momentum of the quarks
(soft)  the energy dependence should be the same as in hadron-hadron
scattering. At higher $k_t$ values one expects the Pomeron to be described by
the two-gluon model, i.e., the energy dependence will be provided by the
square of the gluon structure function of the proton, and consequently a
steeper growth.  In this model the diffractive structure function is given by:
\begin{eqnarray} F_2^{D(3)}   =   F_2^{D(3) ,I} + F_2^{D(3), II} + F_2^{D(3),
III} \,\,, \label{somaf2d} \end{eqnarray} where  the ($I$) and ($II$)
contributions correspond to the production of a quark-antiquark pair and the
production of a quark-antiquark-gluon system with transversely polarized
photon. The third component ($III$) corresponds to the production of a
quark-antiquark pair from a longitudinally polarized photon, which is a
contribution at higher twist (twist-4). Here we desconsider the secundary
reggeon contribution dominant  at low $\beta$. In the leading twist transverse
contribution to $q\bar{q}$ production there is no $\ln(Q^2/Q_0^2)$ enhancement
from the phase-space integral, whereas $q\bar{q}g$ production is of higher
order in $\alpha_s$ and presents an  $\alpha_s\ln (Q^2/Q_0^2)$ dependence. The
longitudinal contribution belongs to higher twist  and the phase-space
integral gives a $\ln (Q^2/Q_0^2)$ enhancement. In a comparison with data, the
transverse $q\bar{q}$, $q\bar{q}g$ production and the longitudinal $q\bar{q}$
production dominate in distinct regions in $\beta$, namely medium, small and
large $\beta$ respectively \cite{BEKW}.  The $\beta$ spectrum and the
$Q^2$-scaling behavior follow from the  evolution of the final state partons,
and are derived from the light-cone wave functions of the incoming photon, 
decoupling from the dynamics inside the Pomeron, while the energy dependence
and the normalizations are  free parameters.

Before to perform the analyzes of the presented models, some comments are in
order. In the first extraction of the $F_2$ slope data at HERA, the so-called
Caldwell plot \cite{caldwell}, the variables $x$ and $Q^2$ were strongly
correlated due to the poor statistics. Since a similar situation should be
present in the first studies of the diffractive slope, in this work
we consider a kinematical constraint  which relates the variables $x$ and
$Q^2$, taken from  the most recent global analyzes of the MRST group
\cite{MRST99}, where the behavior   of the proton structure function slope was
considered. We will  address the behavior of the $ F_2^{D(3)}$ slope
without such constraint in a forthcoming paper. Below, we present our results
for the logarithmic slopes of the diffractive structure function considering
the kinematical constraint.

Starting by the pure CKMT model, we show the dependence on $x_{\pom}$ of the logarithmic
slope at three distinct fixed $\beta$ values in  Fig. \ref{fig1} (a). 
 The slope is ever positive for small
$\beta=0.04$. For medium and high $\beta$ (0.4 and 0.9 values) the
slope is negative for $x_{\pom}<10^{-3}$.  Moreover, the CKMT provides a
transition between positive and negative slope values at $\beta=0.4$. This
behavior is consistent since the Pomeron structure function in this model is
related to the nucleon structure function $F_2$ [Eq. (\ref{deut})], which
presents that feature due to the scaling violation.

The pQCD model provides a quite different result, as presented in the Fig.
\ref{fig1} (b).  The slope is predominantly positive in almost all $\beta$ range, taking negative
values only at $\beta = 0.9$ for the interval $x_{\pom}<0.0004$. A remarkable
feature is the existence of a $\beta$ dependent turn over, which is
shifted to greater $x_{\pom}$ values as $\beta$ increases. The positive
behavior of the slope at low values of $\beta$ is associated to the $q\bar{q}g$
contribution, while for intermediate $\beta$
the $q\bar{q}T$ state dominates, producing an almost constant function on
$Q^2$. The high $\beta$ behavior is consistent with the H1
measurements, in the region $x_{\pom}>10^{-3}$, which prefer a positive slope
in $Q^2$, corresponding to a large $q\bar{q}G$ contribution in this region
\cite{BEKW}.

 For both models the
slope converges to a flat behavior at large values of $x_{\pom}$, with
different behaviors at small $x_{\pom}$ corresponding to low virtualities
($Q^2\le 5$ GeV$^2$). Indeed, the kinematical constraint implies that at
$x_{\pom}=10^{-4}$ we are probing $Q^2 \sim 10^{-3}$, which is  far from the
current  HERA kinematical region.  Confronting the approaches, we conclude
that both models predict a positive slope up to $\beta\sim0.4$, with a steeper
decreasing in CKMT. The  high $\beta$ region discriminates the
behaviors. The pQCD results hold a positive slope, while CKMT produces
negative values. These come from the fact that  the CKMT approach does not
include the $q\bar{q}g$ contribution, which is dominant in this region for
the pQCD model. Since the $Q^2$-behavior in the CKMT is determined by the
$F_2$ scaling violations, then it only includes at most the
$q\bar{q}_{T,\,L}$ contributions. Therefore, the experimental analyzes in this
specific region  of the slope  should clarify the dynamics  in diffractive DIS.

We also present the $Q^2$-slope  as a
function of the variable $\beta$ in Fig. \ref{fig2} for   typical  values
of $x_{\pom}$. Some of the  remarkable features are: (i) the CKMT
 and the pQCD model provide a similar shape (flat behavior)  for the whole
interval of  $\beta$ at  $x_{\pom}\ge
10^{-3}$; (ii)  a noticeable  difference between the  Regge and the pQCD-inspired model in the region of small values of
$x_{\pom}$ ($10^{-4}$), with the prediction of a turn-over at $\beta = 0.1$
 from CKMT while for the pQCD one expects the turn-over at $\beta = 0.5$.
Again, the scaling violations of $F_2$ drive the behavior of the $Q^2$-slope
in CKMT, which implies positive values of the slope at $\beta < 0.5$  and
negative values at larger values. Moreover, this connection implies the large
value of the slope at small $\beta$ and a similar turnover that one found in
the first measurements of inclusive structure function slope \cite{caldwell}.

In  Fig. \ref{fig3} we show the results for $d \ln F_2^D/d
\ln(1/x_{\pom})$ (or shortly, $x_{\pom}$-slope) as a function of the photon virtuality $Q^2$. 
Indeed, this
quantity gives the Pomeron intercept and its behavior describes the energy dependence of the 
diffractive structure function. While the CKMT model predicts a
constant value, without dependence on  $\beta$, the pQCD model presents  a 
 dependence on the $\beta$ value considered. This feature is  associated to  the  distinct energy 
dependence of each  term in Eq. (\ref{somaf2d}), which dominates at specific
regions of the phase space.   A feature in the result is the characteristic
shape of this slope at $\beta = 0.9$, providing a clearly hard intercept. In
fact a dependence on $\beta$ for the Pomeron intercept is expected as shown in
Ref. \cite{BEKW}.    In  principle, the model is only  valid above of the 
starting point $Q_0^2=1\,GeV^2$, however one extrapolated it for lower
virtualities for comparison. For completeness we include the soft Pomeron
intercept (Donnachie-Landshoff) \cite{dola}  in the plot. We verify,
therefore, the evident distinction between the prediction from the CKMT and
pQCD based approaches.

New quantities to distinguish the regimes of QCD have been argued for future
measurements \cite{carta}. The available experimental results seem already
allow to extract information about the slope of the diffractive structure
function, which we propose to study  as a potential source to
discriminate between the hard and soft contribution in diffraction. Considering
two sound  models in the literature, we verify that the results are quite
distinct, allowing to characterize the dynamics through  that quantity.

\section*{Acknowledgments}

This work was partially supported by CNPq and by PRONEX (Programa de Apoio a
N\'ucleos de Excel\^encia), BRAZIL. MVTM acknowledges the DF-UFPel for  their 
warm hospitality. MBGD acknowledges enlightening discussions with Drs. Carlos
Pajares and Alphonse Capella and the hospitality of the Departamento de
F\'{\i}sica de Part\'{\i}culas, U. Santiago de Compostela, where part of this
work was accomplished. VPBG thanks FAPERGS and CNPq for support.

\newpage
\section*{Figure Captions}

\begin{figure}[t]
\vspace{2cm}
\centerline{
\psfig{file=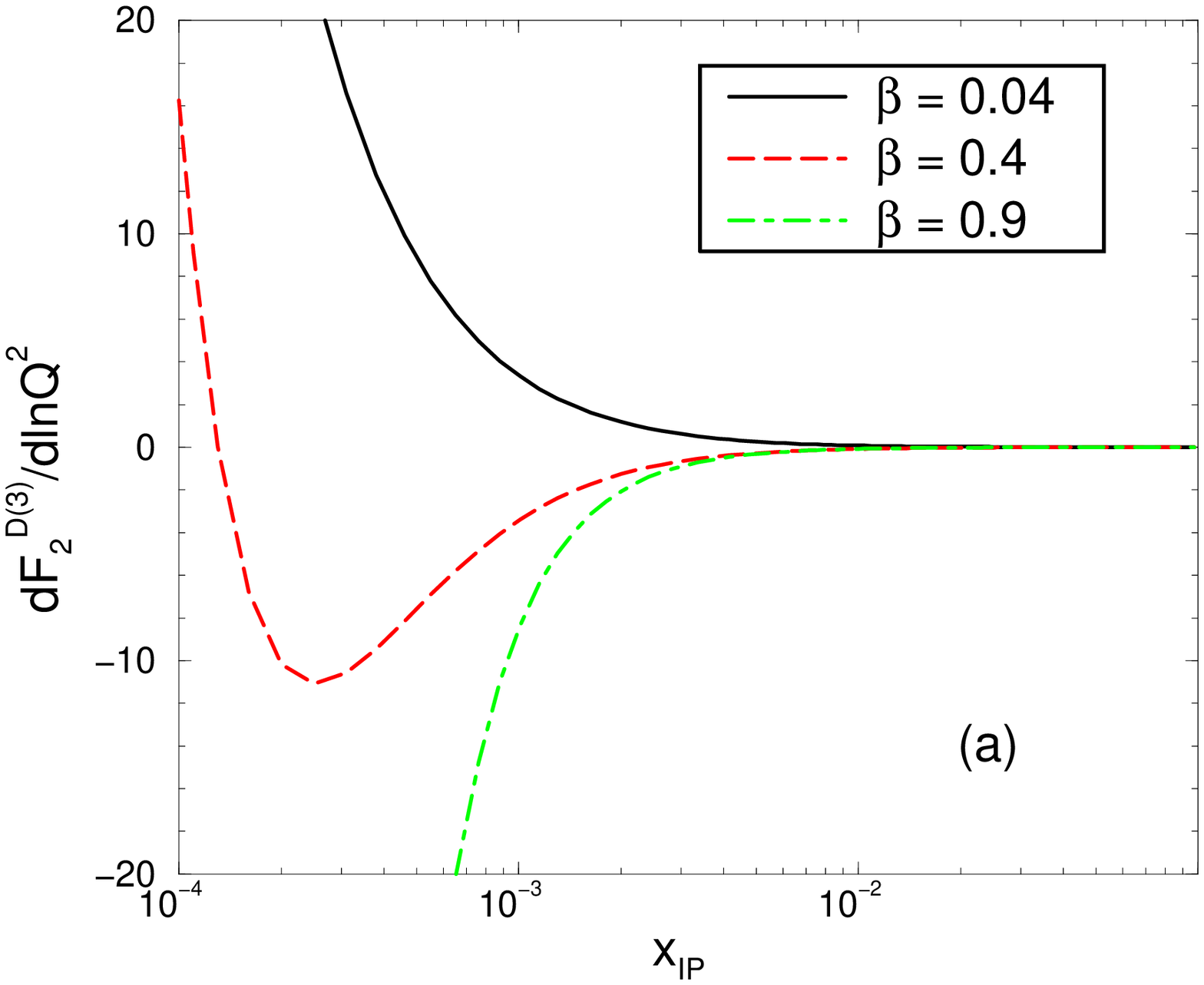,width=80mm} \\ \vspace{1cm}
 \psfig{file=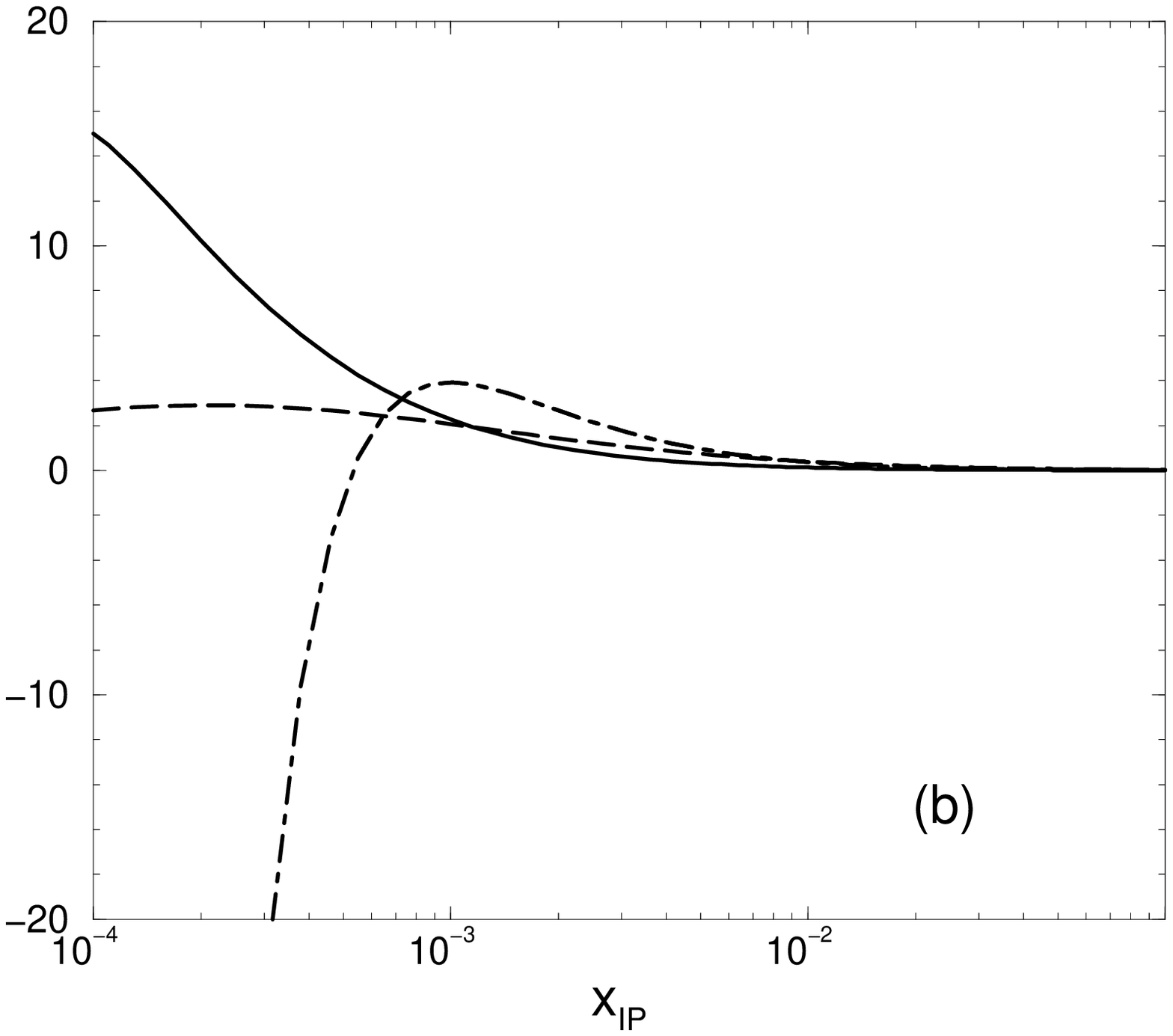,width=80mm}\\}
\vspace{1cm}
\caption{ }
\label{fig1}
\end{figure}

\vspace{1.0cm}
Fig. \ref{fig1}: The $x_{\pom}$ dependence of the $Q^2$-slope at some typical
values of $\beta$ for: (a) the pure CKMT model [9]; (b) the pQCD-inspired
model [11]. Kinematical contraint $Q^2=Q^2(x)$  from the  MRST group [13]  was
used.

\begin{figure}[t]
\vspace{2cm}
\centerline{
\psfig{file=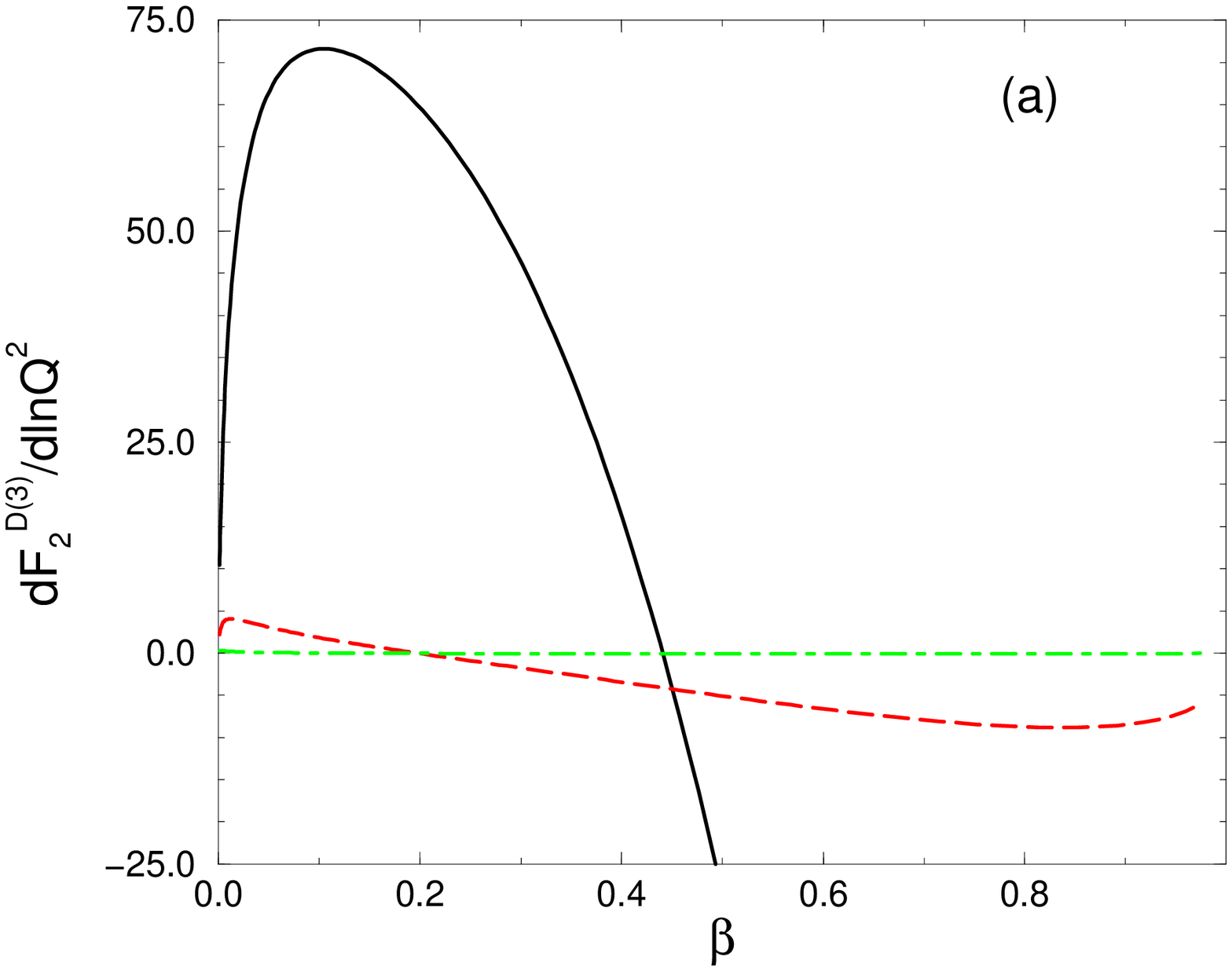,width=80mm} \\ \vspace{1cm}
 \psfig{file=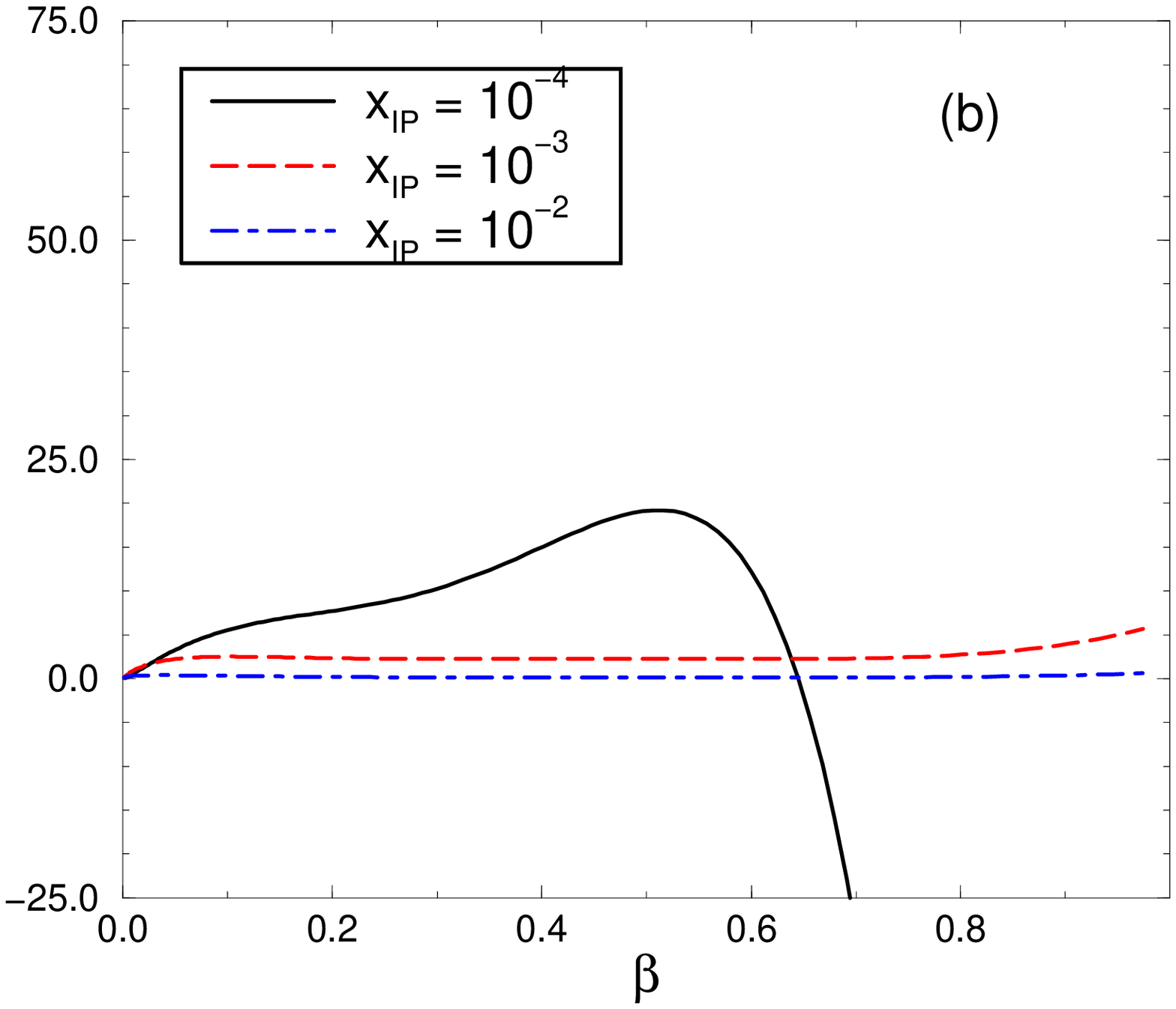,width=80mm}\\ }
\vspace{1cm}
\caption{ }
\label{fig2}
\end{figure}

\vspace{1.0cm}
Fig. \ref{fig2}: The $\beta$ dependence of the $Q^2$-slope at some typical values of 
$x_{\pom}$ for: (a) the pure CKMT model [9]; (b) the pQCD-inspired model
[11]. The kinematical contraint $Q^2=Q^2(x)$  from the  MRST group [13] was
used.

\vspace{1.0cm}
Fig. \ref{fig3}: The $x_{\pom}$-slope versus $Q^2$ for the pQCD approach (BW) [11] and the  
CKMT model [9]. The Donnachie-Landshoff intercept [14] is also depicted.

\newpage

\newpage

\newpage

\begin{figure}[!h]
\centerline{\psfig{file=comparq2.eps,width=100mm}}
\caption{}
\label{fig3}
\end{figure}

\end{document}